Astro 2010
RFI White Paper

# DUAL Gamma-Ray Mission

*Focusing on deep observations of SNe Ia
and wide-field nuclear γ-ray astrophysics.*

Submitted to PPPs: Program Prioritization Panels

**S. Boggs[1], C. Wunderer[1], P. von Ballmoos[2], T. Takahashi[3], N. Gehrels[4], J. Tueller[4], M. Baring[5], J. Beacom[6], R. Diehl[7], J. Greiner[7], E. Grove[8], D. Hartmann[9], M. Hernanz[10], P. Jean[2], N. Johnson[8], G. Kanbach[7], M. Kippen[11], J. Knödlseder[2], M. Leising[9], G. Madejski[12], M. McConnell[13], P. Milne[14], K. Motohide[3], K. Nakazawa[15], U. Oberlack[5], B. Phlips[8], J. Ryan[13], G. Skinner[4], S. Starrfield[16], H. Tajima[12], E. Wulf[8], A. Zoglauer[1], A. Zych[17]**

[1] Space Sciences Laboratory, University of California, Berkeley
[2] Centre d'Etude Spatiale des Rayonnements, Toulouse, France
[3] Institute of Space and Astronautical Science, JAXA, Kanagawa, japan
[4] Astrophysics Science Division, NASA Goddard Space Flight Center
[5] Physics & Astronomy, Rice University
[6] Physics, Astronomy, Ohio State University
[7] Max Planck Institut für extraterrestrische Physik, Garching, Germany
[8] Naval Research Laboratory
[9] Physics & Astronomy, Clemson University
[10] Institut de Ciències de l'Espai (CSIC-IEEC), Barcelona, Spain
[11] Los Alamos National Laboratory
[12] Stanford Linear Accelerator
[13] Space Science Center, University of New Hampshire
[14] Steward Observatory, University of Arizona
[15] Physics, University of Tokyo, Japan
[16] Physics, Arizona State University
[17] Physics, University of California, Riverside

## KEY SCIENCE GOALS

Gamma-rays at MeV energies provide a unique window on the high energy Universe, especially so for nuclear astrophysics. The potential for significant contributions, e.g., to the understanding of SNe Ia as well as the large potential for new discoveries has long been recognized, but technical progress in this challenging energy band has been slow. The groundbreaking discoveries of CGRO's COMPTEL, however, have inspired and driven the development of powerful new instrumentation over the past decade. Novel detector technologies developed in the US and Japan enable compact Compton telescopes, greatly improving the efficiency and field-of-view achievable with this technique successfully employed by COMPTEL. In parallel, Laue diffraction lenses developed in Europe have demonstrated the ability to focus MeV γ-rays over broad energy bands in a telescope configuration. The combination of these two developments, Laue lenses and compact Compton telescopes, has paved the way for the first-ever focusing mission for γ-ray astronomy. In such a configuration, the Compton telescope serves dual roles simultaneously, both as an optimal focal-plane sensor for deep focused observations (with the Compton imaging effectively collimating the focal plane), and as a sensitive wide-field Compton imager in its own right for all-sky surveys and monitoring. This combination addresses the issue of mass and cost inherent in previous Compton-only designs, yet can provide very good sensitivity for deep observations. Our international collaboration is studying this mission with the goal of proposing this as the next γ-ray astrophysics mission.

Gamma-ray astronomy presents an extraordinary scientific potential for the study of the most powerful sources and the most violent events in the Universe. In order to take full advantage of this potential, the next generation of instrumentation for this domain will have to achieve an improvement in sensitivity over present technologies of at least an order of magnitude. The **DUAL** mission concept takes up this challenge in two complementary ways: a very long observation of the entire sky, combined with a large collection area for simultaneous observations of Type Ia SNe. While the Wide-Field Compton Telescope (WCT) accumulates data from the full γ-ray sky (0.1-10 MeV) over the entire mission lifetime, the Laue-Lens Telescope (LLT) focuses on $^{56}$Co emission from SNe Ia (0.8-0.9 MeV), collecting γ-rays from its large area crystal lens onto the WCT. Two separated spacecraft flying in formation will maintain the DUAL payloads at the lens' focal distance.

Focusing instruments have two tremendous advantages: first, the volume of the focal plane detector can be made much smaller than for non-focusing instruments, and second, the residual background, often time-variable, can be measured simultaneously with the source, and can be reliably subtracted. The concept of a Laue γ-ray lens holds the promise of extending focusing capabilities into the MeV range. In order to achieve the ultimate sensitivity for the γ-ray lens mission, the focal plane detector must be designed to match the characteristics of the lens' focal spot. A Compton telescope is a good solution, because the focal plane is intrinsically finely pixellated, optimized for MeV γ-ray detection, and because the direction of incident γ-rays can be determined by the Compton reconstruction to enable discrimination γ-rays coming from the lens ("electronic collimation").

### SUPERNOVAE AND NUCLEOSYNTHESIS

Supernovae synthesize most of the elements heavier than carbon and supply much of the energy input to the interstellar medium. Most of the newly created atoms are indistinguishable from preexisting atoms; however, radioactive species among the ejecta serve as definitive tracers of the nuclear processing. The γ-ray lines from the decay of these nuclei reveal the location of the radioactivity within the ejecta through the time-dependence of the photon escape, and the ejection velocities of various layers through the line Doppler profiles. Therefore spectroscopy and light curve measurements of these γ-ray lines allow direct measurement of the underlying explosion physics and dynamics. The lines from interstellar nuclei paint a unique picture of ongoing nuclear and high-energy processes throughout the Galaxy.

STANDARD CANDLES & ALCHEMISTS

SNe Ia, the thermonuclear explosions of degenerate white dwarfs, are profoundly radioactive events. As much as one-half of the white dwarf mass is fused to $^{56}$Ni ($t_{1/2}$ =6.1d). After a short time, the decays of this nucleus and daughter $^{56}$Co ($t_{1/2}$ =77d) power the entire visible display of the supernova. Most of this power, however, originates in the form of γ-ray lines, some of which begin to escape after several days (Fig. 1a). Thermonuclear supernovae are grand experiments in reactive hydrodynamic flows. Fundamental uncertainties in the combustion physics lead directly to differences in $^{56}$Ni yields and locations[i], which in turn are directly observable with a sensitive γ-ray telescope. Previous attempts to detect $^{56}$Co emission, a primary goal of earlier missions, were unsuccessful due to the instrument sensitivities and supernova distances[ii,iii,iv]. The γ-rays are the most direct diagnostic of the dominant processes in the nuclear burning and explosion.

Fundamental questions about these explosions remain unanswered. We do not understand:
- The physics behind the empirical calibration of their absolute magnitudes that allows them to be used as standard candles for measuring acceleration of the Universe.
- The nature of the progenitor systems[v]– whether the white dwarf companions are normal stars or other white dwarfs.
- How the nuclear flame propagates, how it proceeds as fast as it does, and if, or where, it turns into a shock[vi].

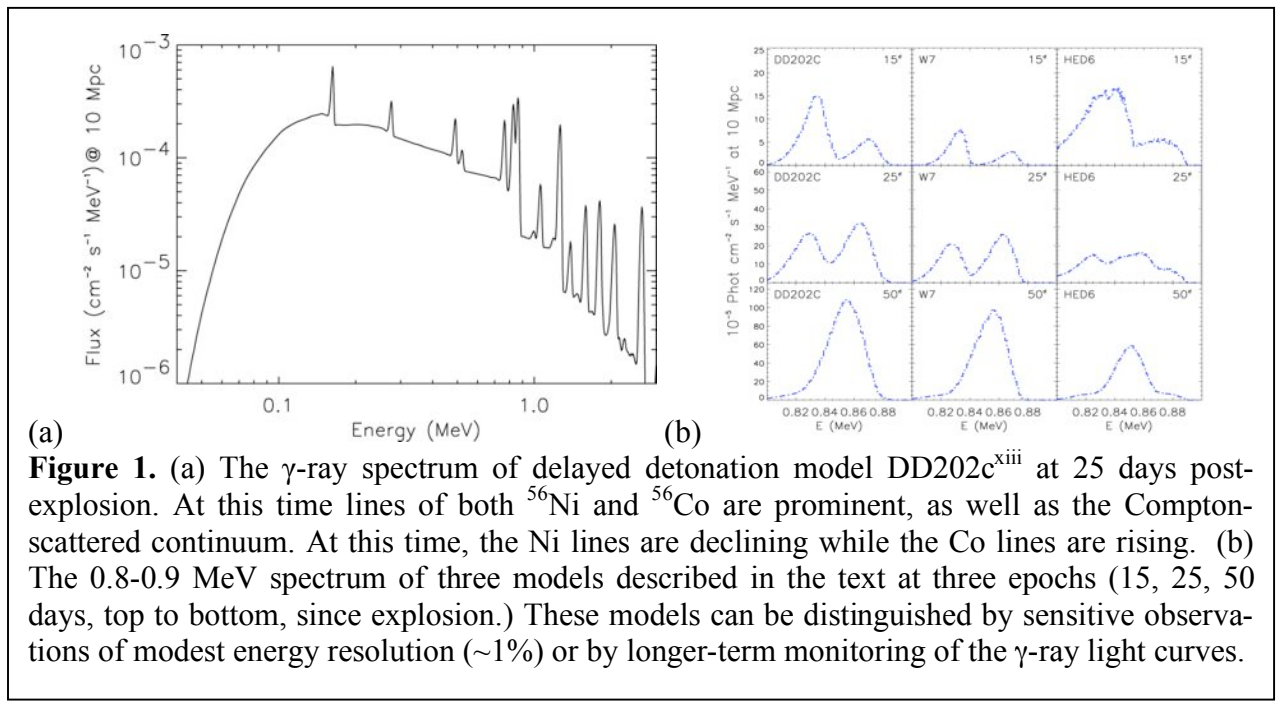

**Figure 1.** (a) The γ-ray spectrum of delayed detonation model DD202c[xiii] at 25 days post-explosion. At this time lines of both $^{56}$Ni and $^{56}$Co are prominent, as well as the Compton-scattered continuum. At this time, the Ni lines are declining while the Co lines are rising. (b) The 0.8-0.9 MeV spectrum of three models described in the text at three epochs (15, 25, 50 days, top to bottom, since explosion.) These models can be distinguished by sensitive observations of modest energy resolution (~1%) or by longer-term monitoring of the γ-ray light curves.

- To what extent instabilities break spherical symmetry, or whether their effects are wiped out by subsequent burning.
- 

Nuclear γ-ray lines from SNe Ia hold the key to solving these puzzles. There are two primary goals of these studies:

1. Standard Candles. Characterize the $^{56}$Ni production distribution for SNe Ia, and correlate with the optical lightcurves to determine the relationship between empirical absolute magnitude corrections and $^{56}$Ni production.

2. Explosion Physics. Clarify the nuclear flame propagation by measuring total and $^{56}$Ni masses and their kinematics for a key handful of SNe Ia, to uniquely distinguish among current models of SNe Ia explosions.

UNDERSTANDING STANDARD CANDLES

Although our understanding of SNe Ia is incomplete, they are used as calibrated standard candles to measure cosmologically significant distances – with dramatic implications[vii,viii]. Though their absolute magnitudes scatter, the Phillips relation is used to calibrate them based on their decline rates. This empirical relation assumes there is a one to one mapping between duration and luminosity[ix]. Whether there is one characteristic of the explosion that determines both is not known[x]. Whether the same relationship should hold exactly to redshifts of unity and beyond, when, for example, the metallicity of the systems was lower, is also not clear[xi]. Complete confidence in the accuracy of using SNe Ia to measure such distances awaits better understanding of the explosions themselves.

Direct correlation between the optical properties and the $^{56}$Ni production – which is principal factor in the optical lightcurve variations – is key to these studies. Other possible causes of the light curve variation can be studied in γ-rays, such as $^{56}$Ni distribution in velocity (through spectroscopy), and total ejecta mass (through light curve monitoring.) Such measurements will directly probe the underlying physical mechanisms driving the variations in visible light. Understanding the explosion mechanism and dynamics from near events will allow much greater effectiveness of the use of SNe Ia at high z, such as with JDEM.

UNCOVERING THE EXPLOSION PHYSICS

Three dominant SNe Ia scenarios have emerged: (1) Single CO white dwarfs grow to the Chandrasekhar mass by accretion; (2) Double white dwarf mergers; and (3) Helium shell detonations triggering thermonuclear runaways in sub-Chandrasekhar mass white dwarfs. Within the generally favored first scenario, the nuclear flame can proceed entirely sub-sonically as a deflagration, or it might accelerate into a detonation in the outer layers of the exploding white dwarf (a delayed detonation). Fig. 1b shows three models for comparison: a Chandrasekhar mass deflagration (W7[xii]), a delayed detonation (DD202C[xiii]) and a sub-Chandrasekhar mass explosion (HED8[xiii]). The γ-ray spectra from merger scenarios are likely intermediate between sub-Chandrasekhar mass models and Chandrasekhar mass models.

To succeed we must clearly discriminate among these models for several SNe Ia per year, and therefore to distances beyond 20 Mpc. With peak line fluxes of (2-5)×10$^{-5}$ (D/10 Mpc)$^{-2}$ cm$^{-2}$ s$^{-1}$, it is clear that a significant improvement in sensitivity over previous and current missions, by a factor 30-50, is required. With such sensitivity, several tens of SNe Ia will be detected per year.

While the combined information from optical and γ-ray studies will be crucial, γ-ray data alone will constrain the models. The differences between the models manifest themselves in both the light curves and the Doppler-broadened profiles of the nuclear emission lines[xiv] (Fig. 1b). These lines are broadened (3–5%), so broad line sensitivity is the most important instrumental requirement; good energy resolution (<1%) would provide valuable additional information.

THE RADIOACTIVE MILKY WAY

Diffuse line emissions from interstellar radionuclides, electron-positron annihilations, and nuclear excitations by accelerated particles afford us the opportunity to study stellar evolution, the ongoing production of the elements, and the most energetic processes throughout the entire Milky Way. The decay of $^{26}$Al shows directly (Fig. 2a) a million years of massive star and supernova activity[xv]. A deep, wide-field survey would enable detailed studies of the production of $^{44}$Ti, $^{26}$Al, and $^{60}$Fe in various types of supernovae. With greatly improved sensitivity and angular resolution, we expect these apparently diffuse emissions to be resolved, at least in part, into hundreds of distinct regions, associations, and individual objects.

POSITRON ASTROPHYSICS

The bright positron annihilation line and triplet-state positronium continuum delineate the escape of positrons from extreme environments over a million years. Interstellar positrons entrained in galactic magnetic fields provide part of the complex 0.511-MeV map (Fig. 2b), which should include individual supernova remnants and stellar and compact object wind nebulae, and possibly the galactic center. With energy resolution of <1% at 0.511 MeV, the conditions in the various annihilation media would be revealed through the line profile and the annihilation physics.

Despite the fact that the electron-positron annihilation is the brightest γ-ray line in the sky, the source of the Galaxy's bulge positrons remains a mystery. Potential sources include LMXBs[xvi], hypernovae from an episode of starburst activity in the bulge[xvii], SNe Ia[xviii,xix], pulsar winds[xx], and annihilation or decay of proposed light dark matter particles[xxi,xxii,xxiii]. Advances will be achieved by producing superior global maps of annihilation radiation and by its detection from individual compact objects and SNRs. As an example, collective study of nearby SNe Ia remnants (e.g. Tycho's SNR, SN 1006, Lupus Loop) will quantify the contribution from type Ia SNe to galactic positrons. Observations of individual objects will offer new insight into the physics of potential sources, and their contribution to the galactic positron budget.

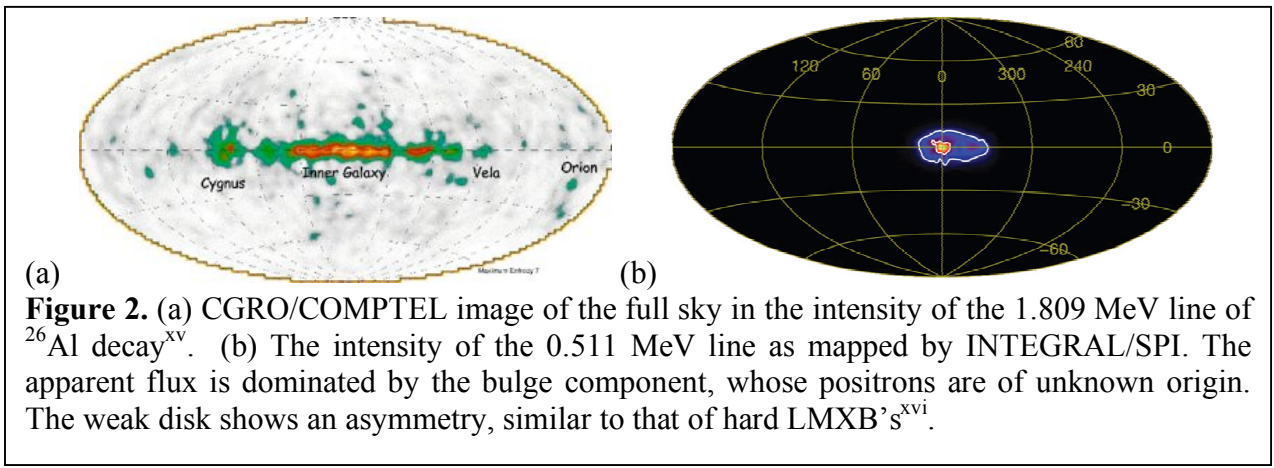

**Figure 2.** (a) CGRO/COMPTEL image of the full sky in the intensity of the 1.809 MeV line of $^{26}$Al decay[xv]. (b) The intensity of the 0.511 MeV line as mapped by INTEGRAL/SPI. The apparent flux is dominated by the bulge component, whose positrons are of unknown origin. The weak disk shows an asymmetry, similar to that of hard LMXB's[xvi].

SCIENCE REQUIREMENTS

A future spectroscopy mission should be able to address emissions with a wide range of angular extent, and with intensities different by several orders of magnitude: The requirements naturally can be divided into two subsets: a requirement for medium-sensitivity large-scale exposures, and very deep pointed observations. This duality is naturally addressed by the DUAL mission concept, which employs a Wide-Field Compton Telescope (WCT) performing deep all-sky surveys in combination with a Laue-Lens Telescope (LLT) that enables simultaneously very deep observations of selected narrow-field targets, utilizing the WCT Compton camera as its focal plane.

WCT: Over its lifetime, the proposed mission should produce all-sky surveys in the energy range of soft gamma rays, i.e. 0.1-10 MeV: mapping out in detail the extended distributions of galactic positron annihilation radiation, and of various long-lived cosmic radioactivities; surveying a large sample of galactic and extragalactic compact sources by characterizing their nonthermal spectra, and study their variability on all timescales; constraining the origin of soft γ-ray cosmic background radiation.

LLT: Simultaneously to the all-sky survey, individual SNe Ia out to distances of 50-80 Mpc will be observed in the 0.847-MeV decay of $^{56}$Co and 0.812-MeV decay of $^{56}$Ni, measuring the intensities, shifts and shapes of the γ-ray lines. An energy resolution of 1% is sufficient to address the DUAL science goals for these Doppler-broadened lines.

| *Parameter* | target (LLT) | all sky (CAST) |
|---|---|---|
| Energy coverage | 0.80-0.90 MeV | 0.1-10 MeV |
| Line sensitivity (ΔE/E=3%) | ~$10^{-6}$ ph cm$^{-2}$ s$^{-1}$ in $10^6$ s | few $10^{-6}$ ph cm$^{-2}$ s$^{-1}$ in 2 y |
| Energy resolution (FWHM) | 1% | 1% |
| FoV | 5' | 2π steradian |
| Angular resolution | 1' | ~1-2° |

Table 1. DUAL mission science requirements

## TECHNICAL OVERVIEW

The DUAL mission is composed of a Wide-field Compton telescope (WCT) and a broad band Laue-Lens Telescope (LLT). Two separated spacecraft flying in formation will maintain the DUAL payloads at the focal length (68 m) by controlling the attitude to within about 1 cm$^3$. Formation flying and orbital constraints are virtually identical to those validated by the CNES/PASO study of the MAX mission concept.

### Wide-Field Compton Telescope (WCT)

Soft γ−ray (0.1-10 MeV) astronomy has undergone a technological revolution through the development of large-volume, high spectral and spatial resolution tracking detectors, developed in large part with NASA investment. Building upon the Compton imaging technique pioneered by COMPTEL on CGRO[xxiv], the modern detectors enable compact telescope designs that improve efficiency by two orders of magnitude, and increase the instrument field-of-view to cover large areas of the sky. Tracking technologies not only provide dramatic improvements in Compton efficiency, but also provide powerful new tools for background rejection.

*Principle:* Compton imaging of γ−ray photons is illustrated in Fig. 3. (For deeper discussions[xxv,xxvi].) An incoming photon of energy $E_\gamma$ undergoes a Compton scatter at a polar angle θ with respect to its initial direction at the position $r_1$, creating a recoil electron of energy $E_1$ that induces the

$$\cos\theta = 1 + \frac{m_e c^2}{E_\gamma} - \frac{m_e c^2}{E_\gamma - E_1}$$

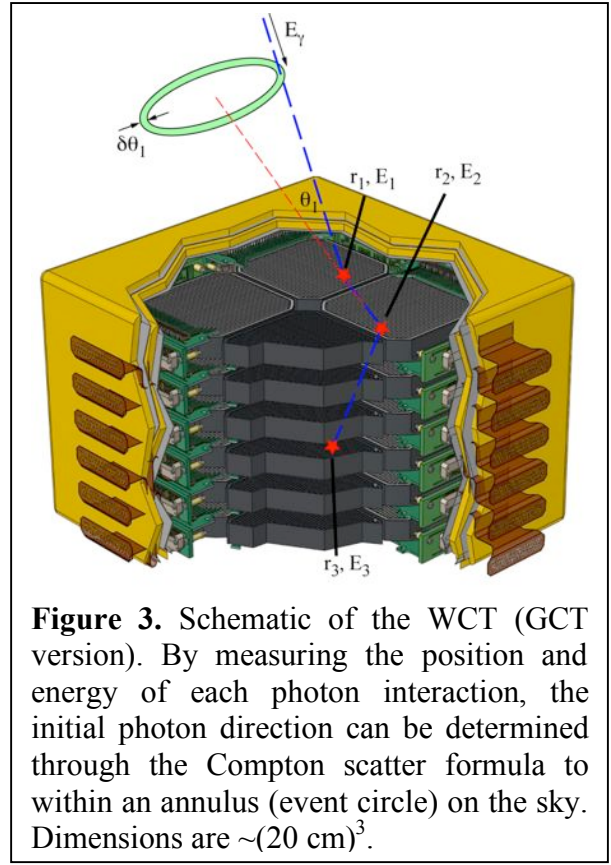

**Figure 3.** Schematic of the WCT (GCT version). By measuring the position and energy of each photon interaction, the initial photon direction can be determined through the Compton scatter formula to within an annulus (event circle) on the sky. Dimensions are ~(20 cm)$^3$.

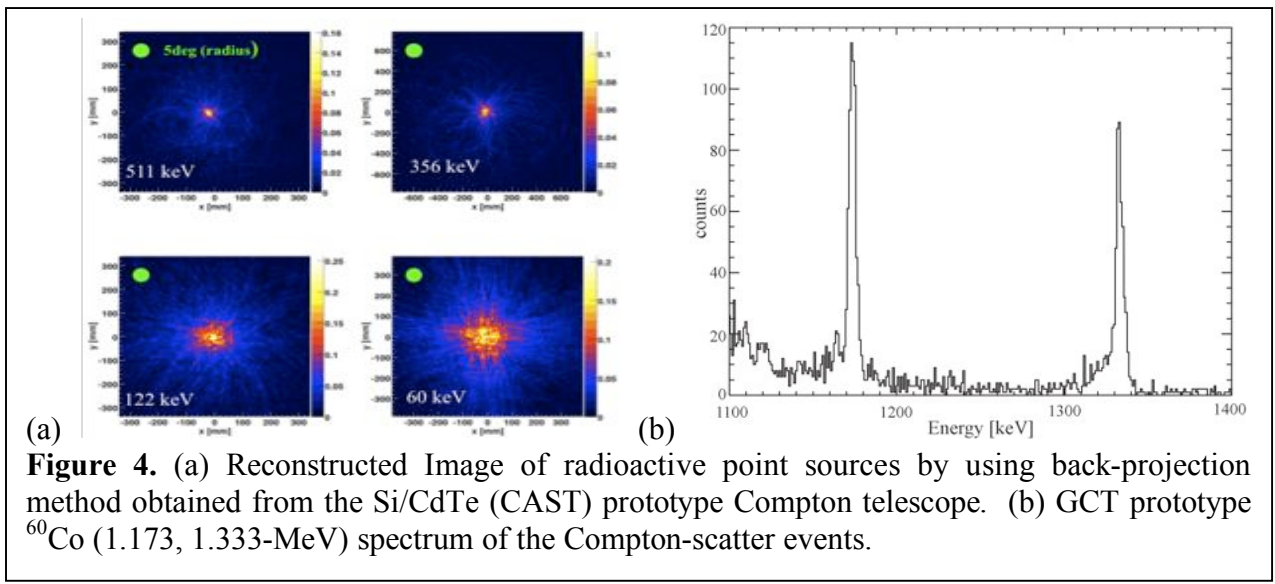

**Figure 4.** (a) Reconstructed Image of radioactive point sources by using back-projection method obtained from the Si/CdTe (CAST) prototype Compton telescope. (b) GCT prototype $^{60}$Co (1.173, 1.333-MeV) spectrum of the Compton-scatter events.

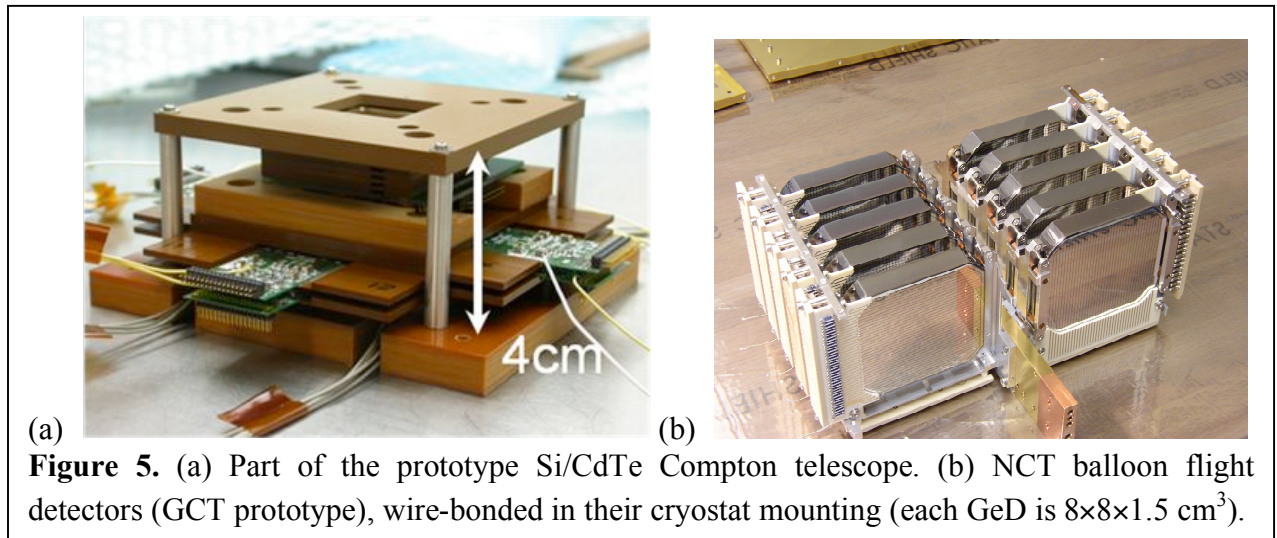

**Figure 5.** (a) Part of the prototype Si/CdTe Compton telescope. (b) NCT balloon flight detectors (GCT prototype), wire-bonded in their cryostat mounting (each GeD is $8\times8\times1.5$ cm$^3$).

signal measured in the detector. The scattered photon then undergoes a series of one or more interactions ($E_i$,$\mathbf{r_i}$), which are also measured. The Compton formula (inset) relates the initial photon direction to the scatter direction (measured direction from $\mathbf{r_1}$ to $\mathbf{r_2}$) and the energies of the incident and scattered γ−rays.

The total energy $E_\gamma$ of the incident γ−ray is determined by summing the measured interaction energies in the case of total absorption (Fig. 4b), which can be uniquely verified for 3 or more interaction sites[xxvi,xxvii,xxviii]. By measuring the position and energy of the photon interactions in the instrument, the event can be reconstructed through use of the Compton formula at each interaction site to determine both the interaction ordering in the detector and the initial photon direction to within an annulus on the sky (Fig. 4a), generally known as the "event circle."

Two instrumental uncertainties contribute to the finite width of the event circle: the uncertainty in θ due to the finite energy resolution, and the uncertainty in the direction between $\mathbf{r_1}$ and $\mathbf{r_2}$ due to the finite spatial resolution. Both of these uncertainties contribute to the uncertainty (effective width) of the event circle δθ. There is also a fundamental limit on the width of the event circle set by Doppler broadening due to Compton scattering on bound electrons, which depends on photon energy and Z of the material (see[xxix] and references therein).

Linearly polarized γ−rays have the highest probability of Compton scattering perpendicular to their polarization vector. This scattering property can be used to measure the intrinsic polarization of radiation from astrophysical sources by measuring a modulation in the distribution of azimuthal scatter angles in the instrument[xxx]. Compact designs maximize the efficiency for detecting photons scattered at θ~90°, which are the most highly modulated, resulting in high sensitivity to polarization.

*WCT Options:* We have several excellent options for WCT, and a DUAL mission could have one or complementary instruments as part of the Compton Telescope payload. In the U.S., researchers have been developing a number of Compton detector technologies for the *Advanced Compton Telescope* Vision Mission[xxxi]. Included in these are the high-resolution germanium detectors developed and flown on the *Nuclear Compton Telescope* (NCT) balloon payload (Fig. 5b). A Ge Compton Telescope (**GCT**) would be based on these technologies. The Japanese have been developing Si/CdTe technologies for the Soft Gamma-ray Detector (SGD) on NeXT (Fig. 5a), and have also developed a mission concept for a follow-on dedicated mission, the Compton

All-Sky Telescope (**CAST**). Both GCT and CAST could serve as an all-sky telescope (Fig. 6), as well as the focal plane detector for the LLT.

**GCT** is a scaled up version of the NCT[xxxii,xxxiii,xxxiv] balloon payload (PI: S. Boggs, UC Berkeley), a NASA suborbital mission developed as a science prototype and technology test bed for the *Advanced Compton Tele-scope*. Both GCT and NCT are wide FOV Compton telescopes utilizing high spatial and spectral resolution germanium cross-strip detectors[xxxv] (GeDs). Each GeD is 15-mm thick, with active area of 54 cm$^2$ (Fig. 5b).

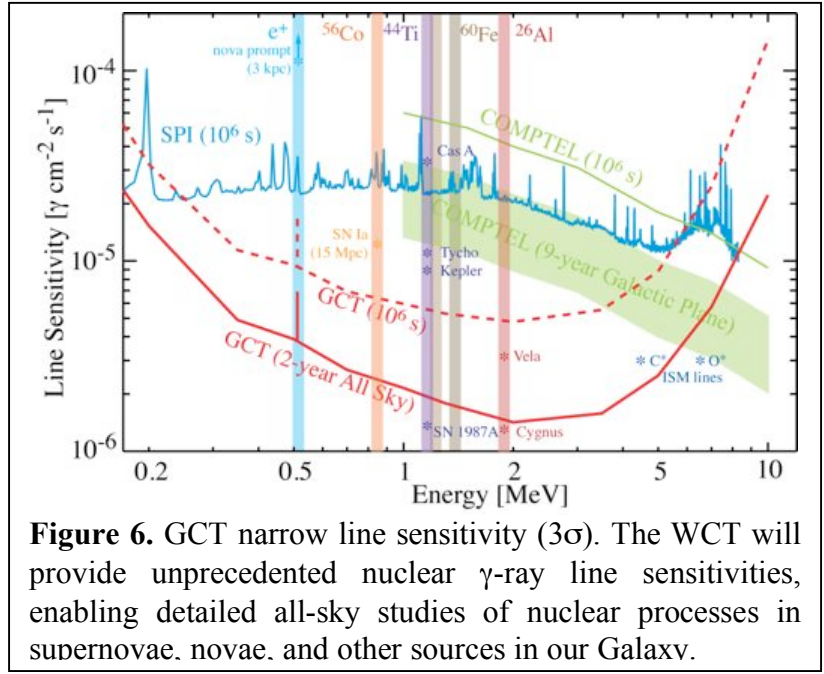

**Figure 6.** GCT narrow line sensitivity (3σ). The WCT will provide unprecedented nuclear γ-ray line sensitivities, enabling detailed all-sky studies of nuclear processes in supernovae, novae, and other sources in our Galaxy.

Photon interaction positions are determined in two dimensions by the active cross strips, and in the third dimension from signal timing difference on the face of the two detectors. For 0.5-mm strip pitch, each GeD would be read out with 320 channels, and achieve 3-D position resolutions of 0.1 mm$^3$. The minimal GCT for achieving the sensitivities would require an array of 24 GeDs, only twice the NCT payload. The array would be enclosed on the sides and bottom by an active 2-cm thick BGO anti-coincidence shield to reduce background. The advantages of GeDs are large volume detectors with uniform response, very high spectral resolution (Fig. 4b), and moderately high Z for photon stopping.

**CAST** is a Japanese small satellite mission concept, presently in a prephase-A study under the Space Science Steering Committee at JAXA (PI: K. Nakazawa, former PI T. Takahashi)[xxxvi], with a target launch window between 2015-2020. CAST is a wide FOV Si/CdTe Compton Telescope based on the SGD (Soft Gamma-ray Detector) on board the 6th Japanese High Energy Astrophysics Mission, called *Astro-H*[xxxvii]. One unit of the telescope combines a stack of ~30 layers of Si PAD detectors and about 8 layers of CdTe PAD detectors with a thickness of 0.7 mm (Fig. 5a). CdTe pad detectors also surround the sides. Abundant R&D on the Si/CdTe Compton camera concept have been conducted by the groups at ISAS and at SLAC[xxxviii,xxxix,xl,xli]. The advantages of the Si/CdTe design are moderate cooling requirements, better intrinsic angular resolution and lower-energy performance from the low-Z Si scatterer (Fig. 5a), and the potential of tracking the recoil electron for higher-energy photons, which can limit the direction of incident γ-rays to an arc, rather than a full cone.

**Laue Lens Telescope**

Simultaneously to the all-sky survey of the WCT, the **Laue-Lens Telescope (LLT)** focuses on a number of selected compact sources, collecting γ-rays from the large collecting area of its crystal diffraction lens onto a very small detector volume. With its outstanding narrow-line sensitivity of the order of $8 \cdot 10^{-7}$ ph·s$^{-1}$·cm$^{-2}$ in the energy band 0.8-0.9 MeV, the focus of the Laue-lens Telescope is on a comprehensive study of $^{56}$Co production and ejection in SNe I.

***Principle:*** The LLT is a broad-band γ-ray lens based on the principle of Laue diffraction of photons in mosaic crystals. (A mosaic crystal consists of little crystallites, slightly misaligned with respect to each other.) Each crystal can be considered as a little mirror that deviates γ-rays through Bragg reflection from the incident beam direction onto a focal spot. Although the Bragg relation ($2d\sin\theta = nhc/E$) implies that only a single energy $E$ (and its multiples) can be diffracted by a given crystal, the crystals' mosaic spread $\Delta\theta$ leads to an energy spread $\Delta E \propto \Delta\theta E^2$ ($d$ is the crystal lattice spacing, $\theta$ the Bragg angle, $n$ the diffraction order, $h$ the Planck constant, $c$ the speed of light, and $E$ the incident photon energy). Placing the crystals on concentric rings around an optical axis, and carefully selecting the inclination angle on each of the rings, allows building up a broad-band γ-ray lens that has continuous energy coverage over a specified band. Since large energies $E_1$ imply smaller diffraction angles $\theta$, crystals diffracting large energies are located on the inner rings of the lens. Conversely, smaller energies $E_2$ imply larger diffraction angles; consequently the corresponding crystals are located on the outer rings (Fig.7).

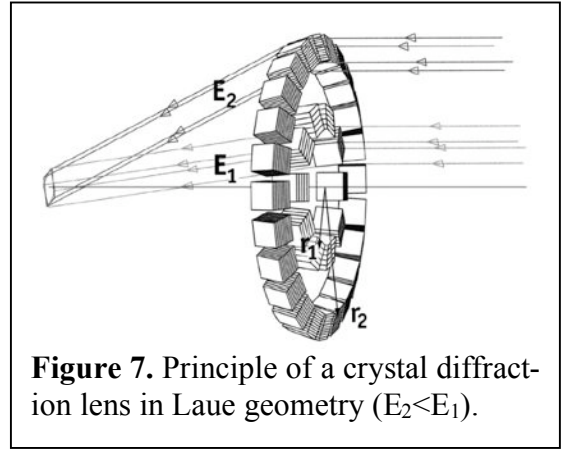

**Figure 7.** Principle of a crystal diffraction lens in Laue geometry ($E_2 < E_1$).

***Model payload for the LLT:*** The most recent R&D activities of our collaboration have confirmed the outstanding performance of Au. The model payload Laue lens is made of 1481 Au and 1129 Cu crystal tiles of 1.5 cm x 1.5 cm each, representing a total geometrical area of 5873 cm² and a total weight of 41 kg. The resulting lens has a focal length of 68 m, with 30" crystal mosaicity, an inner lens radius of 40 cm, and an outer radius of 58 cm. The lens is designed to focus in an energy band from 0.80 to 0.90 MeV, with a total effective area of ~ 500 cm² (Fig. 8) and broad-line sensitivity better than $10^{-6}$ cm$^{-2}$ s$^{-1}$ (Fig. 9).

**Focal Plane Detection:** By providing 3-D localization of the γ-ray interactions, the WCT Compton camera presents a considerable number of advantages as a focal plane detector: besides following the excursions of the focal spot across the detector plane, the WCT allows simultaneous measurement of signal and background. Most importantly, in a system with three-dimensional event localization, a significant background reduction can be achieved by reconstructing the arrival direction of the photons using Compton kinematics. This allows the rejection of photons not coming from the lens direction. In addition, a Compton camera is inherently sensitive to γ-ray polarization (which is unaffected by the lens). Lastly a fine pixellated focal plane allows the limited imaging capabilities of the lens to be used.

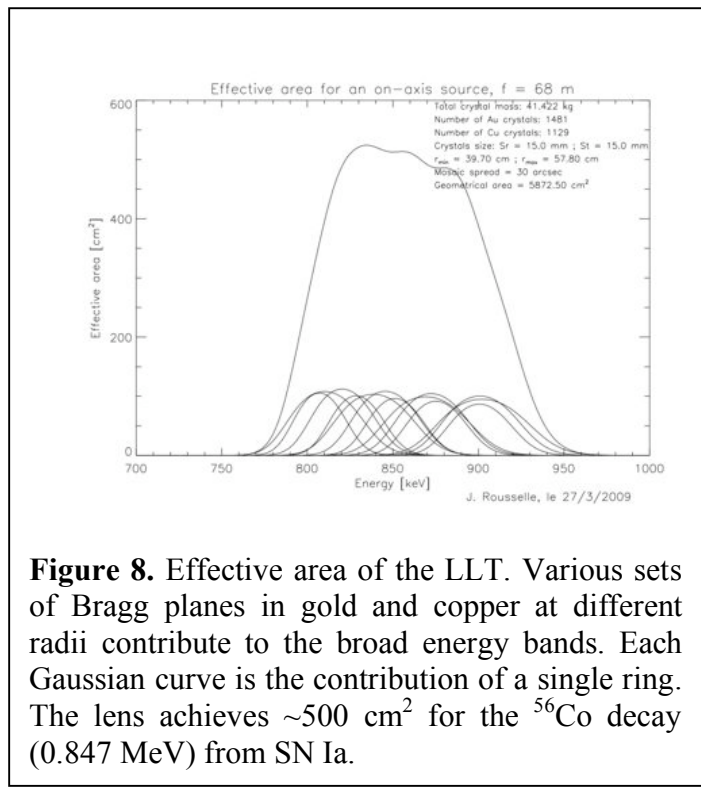

**Figure 8.** Effective area of the LLT. Various sets of Bragg planes in gold and copper at different radii contribute to the broad energy bands. Each Gaussian curve is the contribution of a single ring. The lens achieves ~500 cm² for the $^{56}$Co decay (0.847 MeV) from SN Ia.

**Mission Profile**

DUAL is designed as formation-flying mission composed of two spacecraft: the Optics Spacecraft (OSC) carrying the Laue-Lens Telescope (LLT), and the Detector Spacecraft (DSC) carrying the Wide-field Compton Telescope (WCT). Both satellites will be kept actively in formation at a focal distance of 68 m. The required metrology system for formation flying will also be delivered as a payload item. DUAL would be launched (stacked) by a Delta II into a highly elliptical HEO. Once in final orbit, both satellites will be separated, acquire the Sun, and put into formation. The nominal mission lifetime required to achieve the science objectives is 3 years, with a possible extension necessary for follow-up studies of 1 additional year. The main drivers for the DUAL orbit are (1) moderate propellant consumption for formation keeping and re-orientation to enable a scientific mission duration of at least 3 + 1 years, and (2) an orbit above the proton radiation belts at 10,000 km to avoid the radioactive activation of the detector that would otherwise enhance internal background rates and degrade the instrument's sensitivity. A HEO operational orbit, with perigee between 15,000 and 20,000 km, provides the best conditions for mission maintenance, formation flying control, and ground station link, as well as scientific efficiency. Our mass and power estimates are still somewhat crude. Based on previous studies of a similar instrument, we estimate that the GCT option for the DSC will have a mass of (155 kg science + 122 kg structures + 50 kg hydrazine) = 327 kg, and a total power requirement of 550 W. The 155 kg of science payload includes 78 kg for the active BGO shield. For the OSC, we estimate (70 kg science + 100 kg structures + 300 kg hydrazine) = 470 kg, and a total power requirement of 200 W.

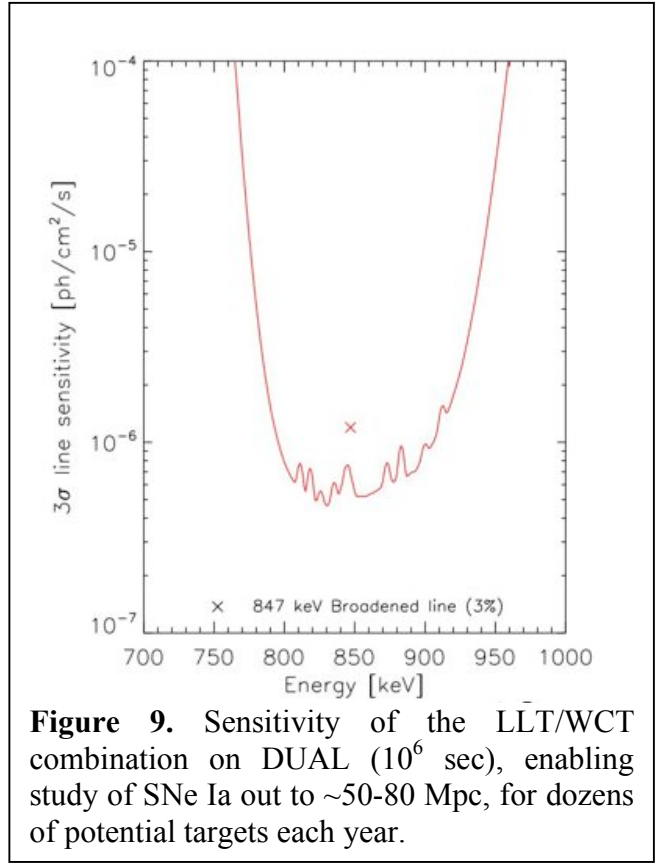

**Figure 9.** Sensitivity of the LLT/WCT combination on DUAL ($10^6$ sec), enabling study of SNe Ia out to ~50-80 Mpc, for dozens of potential targets each year.

**TECHNOLOGY DRIVER I:** *COMPTON TELESCOPE TECHNOLOGIES*

| Item | TRL* | Comments |
|---|---|---|
| *Detector payload* | | |
| Si/CdTe Compton Camera | 4-5 | Prototype detector has been produced and validated in the laboratory. ASICs have been produced and tested for the application in space. |
| Ge Compton Camera | 5-6 | Prototype instrument has flown on the NCT balloon payload. ASICs are currently being produced and tested for the GRIPS payload. |

*CAST:* A team led by T. Takahashi at ISAS and H. Tajima at SLAC are developing this Si/CdTe Compton telescope (see refs 3-8). In addition to the low background achieved by utilizing the Compton kinematics, it features high spectral resolution (2 keV FHWM at 100 keV) and high angular resolution, close to the theoretical limit defined by Doppler broadening. This team has already demonstrated Compton imaging at energies as low as 60 keV.

The proposed detector is based on the Si/CdTe Compton Camera onboard the *Astro-H* mission, planned to be launched in 2013. An engineering model of the Compton Camera will be prepared in 2009. Since the same Camera unit would be used for DUAL, the technical readiness level of CAST will be high. Currently the optimum number of units and optimum configuration to enable the CAST detector to act as both all-sky survey instrument and focal plane instrument is under investigation.

*GCT:* A team led by S. Boggs at UC Berkeley has been developing Ge tracking detectors for the *Nuclear Compton Telescope* (NCT) balloon payload. These detectors have successfully flown on a June 2005 prototype flight, and the instrument is currently (April 2009) in the field for a science flight from Fort Sumner, NM. R. Lin at UC Berkeley has recently been awarded a solar balloon program, GRIPS, utilizing these Ge detector technologies for a solar LDB payload.

Areas for development include: (1) <u>Strip pitch</u>. Detailed studies of Compton event reconstruction in these detectors have shown that event reconstruction (angular resolution, and consequently background rejection and ultimately sensitivity) could be significantly improved by smaller detector voxels. This can be achieved either by reducing strip pitch by a factor of 2-4, a development currently underway in the context of the solar GRIPS balloon, or by inter-strip interpolation. (2) <u>Readout</u>. To be feasible on a space mission, a GCT Ge-strip array requires low-power ASIC readout (on the order of tens of mW/ch). Combining the required low noise, large dynamic range, and fast (<10 ns) triggering, while not exceeding the power envelope, necessitates additional development beyond currently available ASICs. A suitable chip should be developed and tested with Ge-strip detectors. (3) <u>Cooling</u>. While cryocooled Ge detectors have successfully flown in space (RHESSI, INTEGRAL/SPI), the cooling of a GCT array with its significantly increased number of channels and similar Ge mass requires more study. Given the relatively modest size of GCT, several commercial options for cooling are available.

---

* "Technology Readiness level" (TRL) scale 1-9 as used by NASA and ESA. Levels 1 to 4 relate to creative innovate technologies pre or during mission assessment phase. Levels 5 to 9 relate to existing technologies and to missions in definition phase.

### TECHNOLOGY DRIVER II: *LAUE LENS*

| *Item* | *TRL** | *Comments* |
|---|---|---|
| **Lens payload** | | |
| Lens crystals | 4-5 | Ge crystals have been employed and validated during the CLAIRE balloon flight, Au and Cu crystals have been produced and validated in the laboratory. |
| Lens modules | 4 | Bread-board validation in laboratory. |

The proposed lens is based on the experience gained in the Laue lens R&D work conducted during the last decade at CESR (France), UNIFE (Italy), and Argonne National Lab (USA), which culminated so far in the first detection of a γ-ray source (the Crab nebula) using a γ-ray lens during a stratospheric balloon flight (CNES funded CLAIRE project[xlii]). We also benefit from the experience gained during the CNES assessment study of MAX[xliii], an ESA Technical Reference Study on GRL[xliv] and the GRI proposal to ESA in the framework of CV07[xlv]. Furthermore, ASI, CNES and ESA funded R&D work is currently underway to develop space-suitable technology for crystal mounting, control, and accommodation.

**Diffraction Crystals:** The critical performance parameters for a Laue Lens are its effective area and focal spot characteristics (the latter directly relates to the detector volume relevant for the instrumental background) as function of energy. Both parameters are driven by the structure of the individual diffraction crystals. The growing of crystal with the required parameters and homogeneity is making considerable progress in the last years due to dedicated R&D studies by CNES, ESA and ASI. Besides of the "traditional" materials such as Cu, Si and Ge, various high Z crystals (such as Au, Ag, Rh) have recently been measured during runs at the European Synchrotron Facility (ESRF) and a new dedicated reactor-beamline at ILL Grenoble. High-Z mosaic crystals are potentially amongst the most efficient diffraction media. Several of the tested high-Z materials show outstanding performances with reflectivities reaching the theoretical limits for mosaic crystals. Crystal cutting and the determination of the reference planes is still a very time consuming process – optimization of this process is being studied in the framework of the above mentioned R&D programs.

**Crystal lens assembly, lens mechanical and thermal stability:** Crystal mounting, control, and accommodation procedures are currently being developed in dedicated and funded industry studies (CNES contract n°70662/00 with TAS Cannes, ref DCT/SA/AB n°07-3202; and partly in ESTEC contract No. 20357/07/NL/NR) and we expect these issues solved within the next year. In order to achieve a good concentration for the focal spot, each crystal has to be aligned precisely with respect to the lens optical axis. For this purpose, the tolerance on the Bragg angle of each crystal should better than ±10", and the tolerance on the other crystal axes should not exceed ±15'. The dimensional sensitivity of a Laue lens to internal thermal gradients has been investigated by CNES in the framework of the MAX pre phase A study[xlvi]. It has been shown that the thermal internal balance of a comparable Laue lens covered with Multi Layer Insulators (MLI) together with a simple thermal system of thermistors will result in gradients of less than 2°. The resulting thermoelastic deformations are compatible with the individual crystals alignment limit (i.e. 10" of depointing with respect to the lens optical axis).

---

* "Technology Readiness level" (TRL) scale 1-9 as used by NASA and ESA. Levels 1 to 4 relate to creative innovate technologies pre or during mission assessment phase. Levels 5 to 9 relate to existing technologies and to missions in definition phase.

### TECHNOLOGY DRIVER III: *FORMATION FLYING*

| Item | TRL* | Comments |
|---|---|---|
| **Formation Flying** | | |
| formation flying package | 4-5 | virtually identical to the TRL of the Simbol-X formation flying package (validated by the CNES/PASO study of the MAX mission concept) |

While formation flying is not the only option to produce long focal length (booms of extensions up to 100 m have been deployed in space), this technique is to be developed for a number of technically much more challenging missions (LISA, Darwin). With its comparatively modest requirements on station keeping, DUAL can serve as a pathfinder for the more challenging formation flying missions. The master-slave geometry used for DUAL is called "Non-Keplerian Formation Flying" because at least one of the satellites orbit must be continuously controlled. DUAL requires: (1) a lens module able to point and maintain the lens axis to within 15" of the target. As SNe may occur in any direction, there should be as few viewing constraints as possible. (2) a detector module with a sensory and control system enabling it to attain and maintain a precise relative lateral position (± 1 cm) at a focal distance of typically 100 meters (± 0.1 m) and rough target pointing (few degrees). A pre-phase A study of the French Space Agency CNES has demonstrated that the above requirements are realistic and feasible within the framework of a small-satellite mission.

Two types of metrology systems are foreseen for the DUAL formation flying subsystem: RF metrology and optical metrology. Thales Alenia Space (TAS) has designed a possible RF sensor for DUAL. It belongs to the coarse metrology sensors family, providing longitudinal ranging measurements and bearing angles for azimuth and elevation determination. The RF metrology offers also an inter-satellite data-link (ISL), which will be used for HK data transfer from the LLT to WCT. On the LLT spacecraft, a total of 6 antennae are installed, where 3 are facing the WCT in formation flying condition, and 3 are facing opposite to provide 4π coverage for safety reasons. Symmetrically, 6 antennae are installed on the WCT, 3 facing the OSC and 3 on the opposite side. A Hexa-Dimensional Optical Metrology (HDOM) system provides the 3D position and 3D orientation of the WCT relative to the reference frame of the LLT. The HDOM, developed by TAS, is considered for DUAL because of the robustness in luminous environments, allowing for very wide solar aspect angles of ±80° as required for the supernova survey.

---

* "Technology Readiness level" (TRL) scale 1-9 as used by NASA and ESA. Levels 1 to 4 relate to creative innovate technologies pre or during mission assessment phase. Levels 5 to 9 relate to existing technologies and to missions in definition phase.